\documentclass[12pt]{article}
\usepackage{pdproc005} 

\begin{document}

\renewcommand{\thefootnote}{\alph{footnote}}

\title{
 A FLAVOUR-SYMMETRIC PERSPECTIVE ON \\ NEUTRINO MIXING
\footnote{
Presented by W.\ G.\ Scott at ``Neutrino Telescopes'', 
Venice, Italy, 10-13 March 2009.}} 

\author{P. F. HARRISON}

\address{ Department of Physics, University of Warwick, \\
 Coventry, CV4 7AL, UK.\\
 {\rm E-mail: p.f.harrison@warwick.ac.uk}}

  \centerline{\footnotesize and}

\author{W. G. SCOTT}

\address{RAL, Chilton, Didcot, OX11-0QX, UK. \\
 {\rm E-mail: w.g.scott@rl.ac.uk}}

\abstract{A review and consolidation
of some of our more recent publications, 
many with our various collaborators. 
While we cannot resist mentioning Tribimaximal mixing, 
our main theme is Flavour Symmetry,
in particular Flavour-Symmetric Observables, 
scalar (or pseudo-scalar) under $S3_l \times S3_{\nu}$.
Our ``best guess'' for the smallest neutrino mixing angle remains:
$\sin \theta_{13}=\sqrt{2\Delta m^2_{\rm sol}/(3\Delta m^2_{\rm atm})} \simeq 0.13$.}
   
\normalsize\baselineskip=15pt

\section{Introduction}

\noindent

Tri-bimaximal mixing was first put forward in 1999\cite{hps1999}  (10 years ago!)
as a viable alternative to the original ``trimaximal'' ansatz\cite{wolf1978}\cite{cab1978}
on which we were then focussed\cite{hs1994}\cite{hps1995}\cite{hps1996}\cite{hps1997}.
Of course many authors before us had come close 
to proposing tribimaximal mixing,
some very close \cite{wolf1978}\cite{mesh1990}\cite{pak1993}\cite{fz1996}\cite{af1998},
and we can say it is perhaps only that we realised 
the need for a distinct, empirically-based name for this specific mixing pattern
(reflecting its ``bimaximal''\cite{bmx1998}\cite{gold}\cite{glash} 
and ``trimaximal'' character, 
hence ``tri-bimaximal mixing''\cite{hps2002}):
\begin{eqnarray}
     \matrix{  \hspace{0.1cm} \nu_1 \hspace{0.2cm}
               & \hspace{0.4cm} \nu_2 \hspace{0.2cm}
               & \hspace{0.4cm} \nu_3  \hspace{0.2cm} }
                                      \hspace{0.4cm} \nonumber \\
P_{{\rm TBM}} :=(|U_{l \nu}|^2)_{{\rm TBM}} \hspace{2mm} \simeq \hspace{2mm}
\matrix{ e \hspace{0.2cm} \cr
         \mu \hspace{0.2cm} \cr
         \tau \hspace{0.2cm} }
\left( \matrix{ 2/3  &
                      1/3 &
                              0 \cr
                1/6 &
                    1/3 &
                             1/2  \cr
      \hspace{2mm} 1/6 \hspace{2mm} &
         \hspace{2mm}  1/3 \hspace{2mm} &
           \hspace{2mm} 1/2  \hspace{2mm} \cr } \right).
\label{obx}
\end{eqnarray}
In his CERN lecture celebrating 50 years of parity violation, 
T. D. Lee credits us with ``a tremendous achievement''\cite{tdlvid}
(fortunately, elsewhere in his talk, making it clear 
it is the {\em experiments} which are in fact tremendous).
Actually, T. D. Lee's whole lecture\cite{tdlvid} is an
inspiration to anyone working on fermion mixing,
experimenter or theorist alike,
with the CKM and MNS matrices 
cast as ``the cornerstones of particle physics''!

Our original `derivation' of tri-bimaximal mixing\cite{hps2002}
relied on a particular basis, the so-called `circulant basis' (or `cyclic basis') 
where the charged lepton mass matrix (hermitian square) took a cyclically symmetric 
($C3$ invariant), $3 \times 3$ `circulant' form.
Assuming the neutrino mass matrix (hermitian square)
then takes a $C2$-invariant $2 \times 2$ circulant form,
the resulting mixing matrix
(MNS matrix)
$U=U_l^{\dagger}U_{\nu}$ is given~by:
\begin{eqnarray}
    \matrix{  \hspace{0.2cm} \nu_1 \hspace{0.1cm}
               & \hspace{0.2cm} \nu_2 \hspace{0.1cm}
               & \hspace{0.2cm} \nu_3  \hspace{0.2cm} }
                                      \hspace{0.9cm}
\hspace{1.5cm}
    \matrix{  \hspace{0.2cm} \nu_1 \hspace{0.1cm}
               & \hspace{0.2cm} \nu_2 \hspace{0.2cm}
               & \hspace{0.2cm} \nu_3  \hspace{0.2cm} }
                                      \hspace{1.0cm} \nonumber \\
\matrix{ e \hspace{0.2cm} \cr
         \mu \hspace{0.2cm} \cr
         \tau \hspace{0.2cm} }
\left(\matrix{
\frac{1}{\sqrt{3}}
                & \sqrt{\frac{1}{3}} & \frac{1}{\sqrt{3}} \cr
\frac{\omega}{\sqrt{3}}
                & \sqrt{\frac{1}{3}}
                    & \frac{\bar{\omega}}{\sqrt{3}} \cr
\frac{\bar{\omega}}{\sqrt{3}}
             & \sqrt{\frac{1}{3}}
                    & \frac{\omega}{\sqrt{3}} \cr
} \right)
\hspace{0.2cm}
\left(\matrix{
\sqrt{\frac{1}{2}} & 0 & -\sqrt{\frac{1}{2}} \vspace{5pt} \cr
0 & 1 & 0 \vspace{4pt} \cr
\sqrt{\frac{1}{2}} & 0  & \sqrt{\frac{1}{2}} \cr
} \right)
\hspace{0.35cm}
=
\hspace{0.35cm}
\matrix{ e \hspace{0.2cm} \cr
         \mu \hspace{0.2cm} \cr
         \tau \hspace{0.2cm} }
\left(\matrix{
\sqrt{\frac{2}{3}} & \sqrt{\frac{1}{3}} &  0 \cr
-\sqrt{\frac{1}{6}} & \sqrt{\frac{1}{3}} & -\frac{i}{\sqrt{2}} \cr
-\sqrt{\frac{1}{6}}& \sqrt{\frac{1}{3}}   & \frac{i}{\sqrt{2}} \cr
} \right) \hspace{0.80cm}
\label{decomp}
\end{eqnarray}
where the RHS is
the tri-bimaximal form (Eq.~\ref{obx})
in a particular phase convention
($\omega=\exp(i2\pi/3)$ and $\bar{\omega}=\exp(-i2 \pi/3)$
are the complex cube roots of unity).

The above factorisation of tribimaximal mixing, 
into trimaximal and $2 \times 2$ maximal contributions,
has since been exploited by many authors \cite{ma2004} \cite{af2005} \cite{zee2007},
although by no means all derivations invoking
finite groups depend directly on it, see e.g. the $\Delta(27)$ and  $T'$ models\cite{ross}\cite{fram}.
Note that the popular lepton mass matrix $M_l$ of the form\cite{af2005}:
\begin{equation}
M_l =\frac{1}{\sqrt{3}}
\left(\matrix{
m_e & m_{\mu} & m_{\tau} \cr
m_e & m_{\mu}\omega & m_{\tau} \bar{\omega} \cr
m_e & m_{\mu} \bar{\omega} & m_{\tau} \omega
} \right)
\hspace{0.1cm}
\Rightarrow
\hspace{0.1cm}
M_l M_l^{\dagger} =
\left(\matrix{
a & b & b^{*} \cr
b^{*} & a & b \cr
b & b^{*} & a
} \right) \hspace{0.1cm} 
\matrix{
a=\frac{m_e^2}{3} +\frac{m_{\mu}^2}{3}+\frac{m_{\tau}^2}{3} \cr
b=\frac{m_e^2}{3} +\frac{m_{\mu}^2 \omega}{3}
                  +\frac{m_{\tau}^2 \bar{\omega}}{3}
} \hspace{0.1cm}
\label{m2}
\end{equation}
has a `circulant' hermitian square $M_lM_l^{\dagger}$ 
(operating between left-handed fields)
so that our original derivation included this case.
The (charged-lepton) circulant basis 
would indeed seem to be a useful basis
for appreciating possible underlying symmetries.

Of course
the usual (charged-lepton) `flavour basis'
is still the most approriate basis
for understanding neutrino oscillation phenomenology,
matter effects etc.
One may 
ask, then,
how the symmetries are manifested in the charged-lepton flavour basis.
Consider, in particular,
our 2-parameter generalisation\cite{hs2002}\cite{hs2003}
of tribimaximal mixing 
(``$\nu_2$-trimaximal mixing'')
which implements only the constraint that the $\nu_2$ be trimaximally mixed,
interpolating between `tri-chi-maximal' and `tri-phi-maxamal'~mixing\cite{hs2002}:
\begin{equation}
U(\chi,\phi) = \left( \matrix{ 
\sqrt{2/3}c_{\chi} c_{\phi} + i \sqrt{2/3} s_{\chi} s_{\phi} & 
            1/\sqrt{3} & 
                  -\sqrt{2/3}c_{\chi} s_{\phi} - i \sqrt{2/3}s_{\chi} c_{\phi} \cr
        -\frac{c_{\chi} c_{\phi} + i s_{\chi} s_{\phi}}{\sqrt{6}} 
                              - \frac{c_{\chi} s_{\phi} - i s_{\chi} c_{\phi}}{\sqrt{2}} & 
              1/\sqrt{3} & 
                   -\frac{c_{\chi} c_{\phi} - i s_{\chi} s_{\phi}}{\sqrt{2}} 
                              + \frac{c_{\chi} s_{\phi} + i s_{\chi} c_{\phi}}{\sqrt{6}} \cr 
        -\frac{c_{\chi} c_{\phi} + i s_{\chi} s_{\phi}}{\sqrt{6}} 
                                     + \frac{c_{\chi} s_{\phi} - i s_{\chi} c_{\phi}}{\sqrt{2}} & 
              1/\sqrt{3} & \frac{c_{\chi} c_{\phi} - i s_{\chi} s_{\phi}}{\sqrt{2}} 
                          + \frac{c_{\chi} s_{\phi} + i s_{\chi} c_{\phi}}{\sqrt{6}} } \right).
\label{nu2trimax}
\end{equation}
Setting $\phi=0$ gives tri$\chi$-maximal mixing 
and setting $\chi=0$ gives tri-$\phi$maximal mixing
(tri-$\chi$maximal respects $\mu$-$\tau$ reflection symmetry\cite{hsmt},
while tri-$\phi$maximal conserves $CP$).
Clearly a tri-maximal eigenvector (mixing matrix column)
can only result (with the phase choice, Eq.~4)
if the row sums of the neutrino mass matrix ($M_{\nu}$) are all equal,
whereby its hermitian square ($M_{\nu}M_{\nu}^{\dagger}$)
must have all row and column sums equal.
\footnote{Matrices with all row and column sums equal
are clearly expressible as linear combinations of permutation matrices,
and as such play a key role in the representation theory of finite groups
(within a given group and within a given representation 
such `group matrices' \cite{litt}
form a representation of the group `ring' \cite{pass}).
The neutrino mass matrix in the flavour basis
(at least its hermitian square)
is thus empirically an `S3 group matrix'
(in the natural representation of S3) 
whereby we initially dubbed our ansatz (Eq.~4) `S3 group mixng'
(``$\nu_2$-trimaximal mixing'' is simpler and as descriptive).}
%
Thus the neutrino mass matrix in the flavour basis
(in the above phase convention)
is a simple form of `magic-square' \cite{ven2002} \cite{sim2004} \cite{ext2005},
as was emphasised also by Lam \cite{lam2006}.
Our paper with Bjorken\cite{bhs2006} re-parametrises the mixing Eq.~\ref{nu2trimax}
as a function of the complex mixing elememt $U_{e3}$
(it also introduced the notion of a matrix of unitarity-triangle~angles~\cite{hds2009}).

Adding the requirement that the mass matrix be real (no $CP$ violation) 
the mixing Eq.~4 reduces to tri-$\phi$-maximal mixing\cite{hs2002}.
An intriguing construction of this case 
has been given by R.\ Friedberg and T.\ D.\ Lee \cite{leecp} 
based on a kind of `translational invariance' 
in the space of the (neutrino) grassmann fields.
However, their generalisation to the complex case 
seems to involve additional parameters \cite{leecpv}. 
In the (charged-lepton) flavour basis,
our most economical derivation of Eq.~4,
requires only that the neutrino mass matrix 
(hermitian square)
should commute with the `democratic' (mass) matrix 
which is our version of `democracy' symmetry\cite{sim2004}:
\begin{equation}
[M_{\nu}M_{\nu}^{\dagger},D]=0,
\hspace{1.9cm}
{\rm where:}
\hspace{0.3cm}
D =
\left(\matrix{
1 & 1 & 1 \cr
1 & 1 & 1 \cr
1 & 1 & 1
} \right) \hspace{0.2cm} \label{demsym}
\end{equation}
wherby the mixing Eq.~4 then follows,
covering both the real and complex cases. \footnote{
Within a given group and within a given representation, 
class matrices are obtained by summing,
with equal weight, all the group elements within a given class.\cite{litt}
By definition, class matrices commute with all the group elements 
and hence with all group matrices.\cite{litt}
Thus in the case of the natural representaion of $S3$
(see footnote~1) any $S3$ group matrix
commutes with the `democratic' matrix $D$,
which in this case is the only independent, 
non-trivial $S3$ class matrix.\cite{hs2003}\cite{ven2002}}

If we impose instead $\mu$-$\tau$ (reflection\cite{hsmt}) symmetry 
(in the spirit of Lam \cite{lam2001} -
but please note that it \underline{is} a $\mu$-$\tau$ symmetry \cite{hsmt} \cite{kita}
and \underline{not} a 2-3 symmetry \cite{lam2001}!)
the mixing Eq.~4 reduces to the `tri-$\chi$maximal' form.
It should be emphasised that
our `$\mu$-$\tau$ {\em reflection} symmetry' (`mutativity') \cite{hsmt}
includes a complex conjugation
in its definition:
 \begin{equation}
(E^T. M_{\nu}M_{\nu}^{\dagger}.E)^*=M_{\nu}M_{\nu}^{\dagger},
\hspace{1.9cm}
{\rm where:}
\hspace{0.3cm}
E =
\left(\matrix{
1 & 0 & 0 \cr
0 & 0 & 1 \cr
0 & 1 & 0
} \right) \label{mtop}  \hspace{0.2cm} \label{mtsym}
\end{equation}
and thereby predicts a rather strong form of $\mu-\tau$ universality 
$|U_{\mu i}|=|U_{\tau i}| \, \, \, \forall \, \, i=1-3$, 
allowing for non-zero $U_{e3}$ 
and $U_{\mu 3} \ne 1/\sqrt{2}$. 
Thus our symmetry \cite{hsmt} 
is substantively more general than the original (real) form
proposed by Lam\cite{lam2001} and others\cite{kita}.

\section{Jarlskog Invariance and Flavour Symmetry}

Of course all flavour-oscillation observables must be 
basis- and phase-convention independent.
Jarlskog invariance\cite{jarlinv} (also known as weak-basis invariance\cite{braninv})
is readily appreciated working 
in a weak basis 
(where by definition the charged-current weak interaction 
is real, diagonal and universal).
We may then as usual diagonalise the charged-lepton mass matrix
such that the neutrino mass matrix violates charged-lepton flavour,
or, we may (equivalently) choose to diagonlise 
the neutrino mass-matrix instead,
such that the charged-lepton mass matrix violates neutrino ``flavour''
(violation of neutrino ``flavour'' here 
means violation of neutrino {\em mass} as in neutrino decay 
$\nu_3 \rightarrow \nu_2 \gamma$ - we are \underline{not} talking about
old-fashioned 
``neutrino-flavour-eigenstate'' $\nu_e, \nu_{\mu}, \nu_{\tau}$ neutrino flavour!).
Jarlskog invariance is then understood as the freedom 
to `rotate' continuously
in the space spanned by these two extremes
with any unitary transformation
applied to both mass matrices
(the `circulant basis' above
is an example of an `intermediate' basis, 
where neither mass matrix is diagonal).

We are also interested in flavour-symmetry,
and in the use of flavour-symmetric observables in particular.
We will assume here for simplicity
that the freedom to redefine the right-handed fields 
has been used to render the mass matrices themselves hermitian,
and at the same time, we will introduce a useful, more-compact, notation:
\begin{equation}
\vspace{-1mm}
L := M_{l}\hspace{2.0cm} N:=M_{\nu} \label{L=N=}
\vspace{-1mm}
\end{equation}
We may then readily construct the following flavour-symmetric {\em mass} observables:
\begin{eqnarray}
L_1 := {\rm Tr } \, L = m_e+m_{\mu}+m_{\tau} & \hspace{1.0cm} &
N_1 := {\rm Tr} \, N = m_1+m_2+m_3 \label{l1n1} \\
L_2 := {\rm Tr } \, L^2 = m_e^2+m_{\mu}^2+m_{\tau}^2 & \hspace{1.0cm} & 
N_2 := {\rm Tr} \, N^2=m_1^2+m_2^2+m_3^2 \\
L_3 := {\rm Tr } \, L^3 = m_e^3+m_{\mu}^3+m_{\tau}^3 & \hspace{1.0cm} & 
N_3 := {\rm Tr} \, N^3=m_1^3+m_2^3+m_3^3 \label{L1=e+m+t}
\end{eqnarray}
These variables determine the masses through the characteristic equation(s):
\begin{eqnarray}
\lambda_l^3-({\rm Tr} \, L) \lambda_l^2+( {\rm Pr } \, L) \lambda_l -({\rm Det} \, L)=0 \\
\lambda_{\nu}^3-({\rm Tr} \, N) \lambda_{\nu}^2+( {\rm Pr } \, N) \lambda_{\nu} -({\rm Det} \, N)=0
\end{eqnarray}
The coefficients in these equations are themselves flavour-symmetric observables:
\begin{eqnarray}
{\rm Tr } \, L =  m_e+m_{\mu}+m_{\tau}=L_1 \hspace{0.6cm} & \hspace{0.9cm} &
                                    {\rm Tr } \, N  =  m_1+m_{2}+m_{3}=N_1 \hspace{1.2cm} \\
{\rm Pr } \, L  =  m_e m_{\mu}+m_{\mu} m_{\tau}+m_{\tau}m_{e} & \hspace{0.9cm} &
                              {\rm Pr } \, N  =  m_1 m_{2}+m_{2} m_{3}+m_{3}m_{2} \hspace{1.2cm} \nonumber \\
                =  (L_1^2-L_2)/2 \hspace{2.1cm}   & \hspace{0.9cm} & \hspace{1.0cm} =(N_1^2-N_2)/2 \\ 
{\rm Det } \, L  =  m_e m_{\mu} m_{\tau} \hspace{2.5cm} & \hspace{0.9cm} & 
                                       \hspace{-1mm}         {\rm Det } \, N  =  m_1 m_{2} m_{3} \hspace{2mm} \nonumber \\
                 =  (L_1^3-3L_1L_2+2L_3)/6 \hspace{1.5mm} & \hspace{0.9cm} & 
                                         \hspace{1.15cm} =(N_1^3-3N_1N_2+2N_3)/6 \hspace{1.5mm}
\end{eqnarray}
where Pr stands for the sum of the Principal $2\times2$ minors,
invariant under similarity transformations 
as are the trace and determinant
(which are the sum of $1 \times 1$ and $3 \times 3$ principal minors respectively).
Of particular importance are the mass discriminants:
\begin{eqnarray}
L_{\Delta}& :=& \sqrt{L_2^3/2+3L_1^4L_2/2+6L_1L_2L_3-7L_1^2L_2^2/2-3L_3^2-4L_1^3L_3/3-L_1^6/6}
                                                                          \nonumber  \hspace{1.0cm}\\ 
\; & = & \; (m_e-m_{\mu})(m_{\mu}-m_{\tau})(m_{\tau}-m_e). \label{discl} \\
N_{\Delta}& :=& \sqrt{N_2^3/2+3N_1^4N_2/2+6N_1N_2N_3-7N_1^2N_2^2/2-3N_3^2-4N_1^3N_3/3-N_1^6/6}
                                                                          \nonumber  \hspace{1.0cm}\\ 
\; & = & \; (m_1-m_{2})(m_{2}-m_{3})(m_{3}-m_1). \label{discn}
\end{eqnarray} 
which change sign under odd permutations of flavour labels 
(charged-leptons and neutrinos separately).
More precisely, the discriminants have $\bar{1} \times \bar{1}$ symmetry under $S3_l \times S3_{\nu}$,
while all the other flavour-symmetric mass observabes above are $1 \times 1$.

The Jarlskog invariant ${\cal J}$
is the archetypal flavour-symmetric {\em mixing} observable
\begin{equation}
{\cal J} = {\rm Im} \, \Pi_{l \nu}=-i\frac{{\rm Det} \, [L,N]}{2 L_{\Delta}N_{\Delta}}
\label{jinv}
\end{equation}
measuring the violation of $CP$ symmetry,
with no reference to particular flavour labels (Eq.~42 defines $\Pi_{l \nu}$). 
That ${\cal J}$ has $\bar{1} \times \bar{1}$ symmetry under $S3_l \times S3_{\nu}$
follows by virtue of the product of mass discriminants 
$L_{\Delta}N_{\Delta}$ appearing in the denominator of Eq.~\ref{jinv}.

\section{Six New Flavour-Symmetric Mixing Observables}

In terms of the matrix $T$ (see Eq.~\ref{tq}) of traces 
of \underline{anti}-commutators (c.f\ Eq.~\ref{jinv}) 
we define:
\begin{equation}
{\cal F}:= {\rm Det} \, P= \frac{{\rm Det} \, T}{L_{\Delta} N_{\Delta}}
=\frac{{\rm Det} \,  ({\rm Tr} \, \{L^m,N^n\})}{2 L_{\Delta} N_{\Delta}} 
\end{equation}
Somewhat as ${\cal J}=0$ protects a neutrino source 
against matter vs.\ anti-matter analysis,
so ${\cal F}=0$ protects against flavour analysis
(since the transition-probability matrix $\cal{P}$ cannot be inverted if ${\rm Det} \, P=0$). 
${\cal F}$ (like ${\cal J}$) is ``odd-odd'' $(\bar{1} \times \bar{1})$  under $S3_l \times S3_{\nu}$.

We may define all our new mixing observables\cite{hrs2007}
as homogeneous polynomials in $w,x,y,z$, 
parameterising
the $P$-matrix in terms  
of deviations from trimaximal mixing:
\begin{eqnarray}
P \, := \, (|U|^2) =:  \hspace{2mm}
\left( \matrix{ 1/3-w-x  &
                      1/3+w &
                              1/3+x \cr
                1/3-y-z &
                    1/3+y &
                             1/3+z  \cr
      \hspace{2mm} 1/3+w+x+y+z \hspace{2mm} &
         \hspace{2mm}  1/3-w-y \hspace{2mm} &
           \hspace{2mm} 1/3-x-z  \hspace{2mm} \cr } \right).
\label{P=wxyz}
\end{eqnarray}
The polynomials are then chosen so as to have
definite symmetry under $S3_l \times S3_{\nu}$: 
\begin{eqnarray}
{\cal G}:=(1 \times 1)^{(2)} & = & 2(w^2 + x^2 + y^2 + z^2 + w x + w y + x z + y z) + (w z + x y) \label{g=wxyz} \\
{\cal F}:=(\bar{1} \times \bar{1})^{(2)} & = & 3(w z - x y) \\
{\cal C}:=(1 \times 1)^{(3)}& = & 9(x y z + w y z + w x z + w x y) \nonumber \\ 
                     &   & + 9/2(x y (x + y) + w z(w + z)) \\
{\cal A}:=(\bar{1} \times \bar{1})^{(3)}& = & 2(w^3 - x^3 - y^3 + z^3) \nonumber \\ 
  &    & + 3(wx(w - x) + wy(w - y) + yz(z - y) + xz(z - x)) \nonumber \\ 
  &    &    + 3( xy(w + z) - wz(x + y)) \nonumber \\
  &    &         + 3/2(wz(w + z) - xy(x + y)) \label{a=wxyz} \\
{\cal B}:=(\bar{1} \times 1)^{(3)} & = & 3\sqrt{3}[(w^2y + w y^2 - x^2z - z^2x + w x y + w y z - x y z - w x z) \nonumber \\ 
               & & + 1/2(w^2z - w z^2 + x y^2 - x^2y)] \\
{\cal D}:=(1 \times \bar{1})^{(3)} & = & 3\sqrt{3}[(w^2x + w x^2 - y^2z - z^2y + w x y + w x z - x y z - w y z) \nonumber \\ 
               & & + 1/2(w^2z - w z^2 + y x^2 - y^2x)] 
\end{eqnarray}
Our polynomials are all $C3_l \times C3_{\nu}$ invariant
and form a natural basis of the $C3 \times C3$ invariant polynomial ring \cite{db}.
They are ``plaquette invariant'' in that
the definitions above are independent of 
which ``plaquette'' is chosen to define $w,x,y,z$ in Eq.~\ref{P=wxyz}.

With only four parameters needed to fix the mixing magnitudes,
not all of our six flavour-symmetric mixing observables (FSMOs) 
can be independent. We have:
\begin{eqnarray}
{\cal A}^2+{\cal B}^2+{\cal C}^2+{\cal D}^2 & = & {\cal  G}({\cal F}^2+3{\cal G}^2)/2 \\
2({\cal AC}+{\cal BD}) & = & {\cal  F}({\cal G}^2+3{\cal F}^2)/2
\end{eqnarray}
constituting the two constraints needed 
to reduce six to four. 
Of course neither can our variables be entirely 
independent of Jarlskog ${\cal J}$ and we have also:
\begin{equation}
{\cal J}^2=1/108-{\cal G}/18+2{\cal C}/27-{\cal F}^2/36
\end{equation}

\subsection{Expression in terms Mass Matrices}

All our variables can of course be readily written in terms of traces:
\begin{eqnarray}
2{\cal G}-1 & := & {\rm Tr} \, P^T.P ={\rm Tr } \, T^T.L_G.T.N_G \label{gtrace} \\
6{\cal F} & := &{\rm Tr} \, P^T.\epsilon.P.\epsilon^T= {\rm Tr} \, T^T.L_F.T.N_F^T \\
2{\cal C}/3-{\cal G}/2-1/6 & := & {\rm Tr } \, P^T.K={\rm Tr} \, T^T.L_C.Q.N_C^T \\
2{\cal A}-2{\cal F} & := & {\rm Tr} \, P^T.\epsilon.K. \epsilon^T={\rm Tr} \, T^T.L_A.Q.N_A^T \label{atrace} \\
2{\cal B}/\sqrt{3} & := & {\rm Tr } \, P^T. \epsilon.K= {\rm Tr} \, T^T.L_A.Q.N_C^T \\
2{\cal D}/\sqrt{3} & := & {\rm Tr} \, P^T.K.\epsilon^T={\rm Tr} \, T^T. L_C.Q.N_A^T 
\end{eqnarray}
where the $K$-matrix\cite{pkq}
is the real part of the mixing-matrix ``plaquette products''
(Eq.~\ref{Pilnu})
with $\epsilon$ 
the totally anti-symmetric $3 \times 3$ ``epsilon'' matrix
\cite{ven2002}\cite{sim2004} (Eq.~\ref{eps}).

In terms of anti-commutators ($A_{mn}$) and commutators ($C_{mn}$) of mass matrices:
\begin{equation}
A_{mn}=\{L^m,N^n\} \hspace{1.0cm} C_{mn}=-i[L^m,N^n]
\end{equation}
we define the $T$-matrix\cite{pkq} and $Q$-matrix\cite{pkq} as traces of 
anti-commutators (i.e.\ of products) and quadratic (products of) commutators respectively: 
\begin{equation}
T  =  \frac{1}{2} 
\left( \matrix{{\rm Tr} \, A_{00} & {\rm Tr} \, A_{01} & {\rm Tr} \, A_{02} \cr
                  {\rm Tr} \, A_{10} & {\rm Tr} \, A_{11} & {\rm Tr} \, A_{12} \cr
                   {\rm Tr} \, A_{20} & {\rm Tr} \, A_{2,1} & {\rm Tr} \, A_{22} } \right) ;
 \hspace{0.3cm}
Q   =  \frac{1}{2}
          \left( \matrix { {\rm Tr} \, C_{11}^2  & {\rm Tr} \, C_{11}C_{12} & {\rm Tr} \, C_{12}^2  \cr
                 {\rm Tr} \, C_{11}C_{21} &  {\rm Tr} \, C_{11}C_{22} & {\rm Tr} \, C_{12}C_{22}  \cr
                 {\rm Tr} \, C_{21}^2 & {\rm Tr} \, C_{21}C_{22} &  {\rm Tr} \, C_{22}^2 } \right) 
\hspace{-4mm} \label{tq}
\end{equation}
related to the $P$ and $K$ matrices (respectively) by simple mass-moment transforms~\cite{pkq}.

The matrices $L_G, L_F, L_C, L_A$ and $N_G, N_F, N_C, N_A$ 
(Eqs.~\ref{gtrace}-\ref{atrace}) are just 
functions of our flavour-symmetric mass observables (Eqs.~\ref{l1n1}-\ref{L1=e+m+t}) for example:
\begin{eqnarray}
L_G=\left( \matrix{L_0 & L_1 & L_2 \cr
                   L_1 & L_2 & L_3 \cr
                   L_2 & L_3 & L_4 } \right)^{-1} \hspace{0.8cm}
L_F=\frac{1}{L_{\Delta}} \left( \matrix{0 & L_2 & -L_1 \cr
                   -L_2 & 0 & L_0 \cr
                   L_2 & -L_0 & 0 } \right) \\
L_0=3; \hspace{0.5cm} L_4=L_1^4/6+L_1L_3/3+L_2^2/2-L_1^2L_2; 
\hspace{0.5cm} L_{\Delta}={\rm Det} \, L_G^{-1}. \nonumber
\end{eqnarray}
The explicit form of the matrix $L_C$
is somewhat less succinct:
\begin{eqnarray}
& &  L_C   =  \frac{1}{L_{\Delta}^2}  
\left( \matrix{ L_C(1,1) &  (L_1^2 - L_2)/2 & (L_1^2 - L_2)/2 \cr
              -(L_1^3 - L_3) &  3L_1^2 - L_2 & -2L1 \cr
            (3L_1^2 - L_2)/2 &  -4L_1 & L_0 } \right) \hspace{0.5cm} \\
& & L_0  =  3; \hspace{0.5cm}  L_C(1, 1) = L_3 L_1/3 + L_2^2/4 - L_2 L_1^2 + 5 L_1^4/12 \nonumber
\end{eqnarray}
while the $L_A$ matrix is just a little too lengthy to usefully display here.
Obviously, identical expressions (with $L \rightarrow N$) obtain for the corresponding neutrino matrices.

Remarkably our $(\bar{1} \times \bar{1})^{(2)}$ mixing observable ${\cal F}$ 
is not only expressible in terms of the $T$-matrix
but also in terms of the matrix of cubic (products of) commutators\cite{hawaii}: 
\begin{equation}
{\cal F}= \frac{{\rm Det} \, T}{L_{\Delta}N_{\Delta}}
       =\frac{{\rm Det} \, C^{(3)}}{L_{\Delta}N_{\Delta} {\rm Det}^3 \, C_{}} \hspace{0.6cm}
C^{(3)}:= \left( \matrix { {\rm Tr} \, C_{11}^3  & {\rm Tr} \, C_{11}^2C_{12} & {\rm Tr} \, C_{11}C_{12}^2  \cr
                 {\rm Tr} \, C_{11}^2C_{21} &  {\rm Tr} \, C_{11}^2C_{22} & {\rm Tr} \, C_{21}C_{12}^2  \cr
                 {\rm Tr} \, C_{11} C_{21}^2 & {\rm Tr} \, C_{12}C_{21}^2 &  {\rm Tr} \, C_{11}C_{22}^2 } \right)
\end{equation}
where $C:=C_{11}:=-i[L,N]$ is the usual Jarlskog commutator (${\rm Det} \, C_{}={\rm Tr} \, C_{}^3/3$).

\subsection{Interpretation and Flavour-Symmetric Mixing Constraints}

Our $(1 \times 1)^{(2)}$ variable ${\cal G}$ can obviously be related 
to the flavour-averaged asymptotic \cite{pkq} survival probability (via ${\rm Tr} \, P^T.P$). 
Interestingly (and perhaps less obviously)
both ${\cal G}$ and ${\cal C}$ can be related to certain flavour-summed loop amplitudes as follows:
%
\begin{center}
\begin{picture}(100,120)
\thicklines
\put (-100.0,30.0){\line(0,1){50.0}}
\put (-100.0,30.0){\line(-1,-1){10.0}}
\put (-120.0,10.0){\line(-1,-1){10.0}}
\put (-100.0,30.0){\line(1,0){50.0}}
\put (-50.0,30.0){\line(1,-1){10.0}}
\put (-30.0,10.0){\line(1,-1){10.0}}
\put (-50.0,30.0){\line(0,1){50.0}}
\put (-50.0,80.0){\line(1,1){10.0}}
\put (-30.0,100.0){\line(1,1){10.0}}
\put (-100.0,80.0){\line(1,0){50.0}}
\put (-100.0,80.0){\line(-1,1){10.0}}
\put (-120.0,100.0){\line(-1,1){10.0}}
%
%
\put (100.0,40.0){\line(0,1){30.0}}
\put (100.0,40.0){\line(-2,-1){10.0}}
\put (80.0,30.0){\line(-2,-1){10.0}}
\put (100.0,40.0){\line(2,-1){30.0}}
\put (130.0,100.0){\line(0,1){10.0}}
\put (130.0,85.0){\line(0,1){7.0}}
\put (100.0,70.0){\line(2,1){30.0}}
\put (100.0,70.0){\line(-2,1){10.0}}
\put (80.0,80.0){\line(-2,1){10.0}}
\put (160.0,70.0){\line(-2,1){30.0}}
\put (160.0,70.0){\line(2,1){10.0}}
\put (180.0,80.0){\line(2,1){10.0}}
\put (160.0,70.0){\line(0,-1){30.0}}
\put (160.0,40.0){\line(2,-1){10.0}}
\put (180.0,30.0){\line(2,-1){10.0}}
\put (160.0,40.0){\line(-2,-1){30.0}}
\put (130.0,25.0){\line(0,-1){7.0}}
\put (130.0,10.0){\line(0,-1){10.0}}
\end{picture}
\end{center}
\begin{equation}
\sum_{l, \nu} \Pi_{l \nu}=({\cal G}-1)/2 +9i{\cal J} \hspace{2mm}
\hspace{2.0cm} \sum_{l \pm \nu} \Omega_{l\pm \nu}=2/9{\cal C}-1/3{\cal G}+1/9 \hspace{5mm} 
\end{equation}
(we focus purely on the flavour structure here, 
in the limit that all masses are zero).
The individual plaquettes (4-plaquette) and hexaplaquettes (6-plaquette) contribute:
\begin{eqnarray}
\Pi_{l \nu} & := & U_{l-1, \, \nu-1}U^*_{l-1, \, \nu+1}U_{l+1, \, \nu+1}U^*_{l+1, \, \nu-1};
\hspace{0.5cm} K_{l\nu}:={\rm Re } \, \Pi_{l\nu} \hspace{0.5cm} 
{\rm mod} \, 3 \hspace{2mm} \label{Pilnu} \\
\Omega_{l \mp \nu}& := & U_{l-1, \, \pm \nu} U^*_{l-1, \, \pm \nu+1} 
                   U_{l, \, \pm \nu+1} U^*_{l, \, \pm \nu-1} 
                   U_{l+1, \, \pm \nu-1} U^*_{l+1, \, \pm \nu}
\hspace{0.9cm} {\rm mod} \, 3 \hspace{2mm}
\end{eqnarray}
where indices are to be interpreted mod 3, 
i.e.\ $e+1=\mu$, $\mu+1=\tau$, $\tau+1=e$ etc.

Most importantly, flavour-symmetric mixing variables 
can be used to specify (up to permutations)
various suggestive and often phenomenologically relevant 
mixing ansatze with their constraints/symmetries, for example:
\begin{eqnarray}
 {\rm ``Democracy \hspace{1mm} symmetry''} & \Leftrightarrow &  
{\cal F}=0 \hspace{0.9cm} {\cal C}=0 \label{fcdem} \\
{\rm ``\mu-\tau \hspace{1mm} refl.\ symmetry''}   & \Leftrightarrow &  
{\cal F}=0  \hspace{0.9cm} {\cal A}=0 \label{famut} \\
{\rm ``Tribimaximal \hspace{1mm} Mixing''}   & \Leftrightarrow &  
{\cal F}= {\cal C} = {\cal A}= {\cal J}=0 \label{fgcatbm}
\end{eqnarray}
Comparing Eq.~\ref{fcdem} and Eq.~\ref{famut} 
one notices 
a kind of ``duality'' 
between 
the democracy and $\mu-\tau$-reflection symmetries,
with ${\cal C}$ and ${\cal A}$ interchanging roles between the two cases.
The condition for tribimaximal mixing, Eq.~\ref{fgcatbm}, 
is given in terms of the Jarlskog variable \cite{jarlinv}
equivalently setting ${\cal J}=0$ for simplicity,
in place of ${\cal G}=1/6$.

Some further examples 
of ansatze and flavour-symmetric constraints are given in Table~1.
Our variables are normalised such that their maximum values are unity,
corresponding to the case of ``no mixing'' as in the first line of Table~1. 
\begin{table}[h]
  \small
  \begin{tabular}{||l|l|l|l||}\hline\hline
  {} &{} &{} &{}\\
  Mixing ansatz & \hspace{1mm} ${\cal F} \hspace{5mm} {\cal G} \hspace{5mm} {\cal C} \hspace{5mm} {\cal A}$ \hspace{1mm}
                & Symmetries & \hspace{0mm} $18{\cal J}^2$ \hspace{1mm} ${\cal B}$ \hspace{3mm} ${\cal D}$ \hspace{1mm} \\
  {} &{} &{} &{}\\
  \hline
  No Mixing \rule{0mm}{5mm}& \hspace{1mm} $1$ \hspace{5mm} $1$ \hspace{5mm} $1$ \hspace{5mm} $1$ \hspace{1mm}
                & $CP$ & \hspace{2mm} $0$ \hspace{5mm} 0 \hspace{5mm} 0 \hspace{1mm}  \\
  Tribimaximal Mix.$^*$ \rule{0mm}{4mm}& \hspace{1mm} $0$ \hspace{5mm} $\frac{1}{6}$ \hspace{5mm} $0$ \hspace{5mm} $0$ \hspace{1mm}
                & Dem., $\mu$-$\tau$, $CP$ & \hspace{2mm} $0$ \hspace{5mm} $0$ \hspace{2mm} $\frac{1}{12\sqrt{3}}$ \hspace{1mm} \\ 
  Trimaximal Mix. \rule{0mm}{4mm}& \hspace{1mm} $0$ \hspace{5mm} $0$ \hspace{5mm} $0$ \hspace{5mm} $0$ \hspace{1mm}
                & Dem., $\mu$-$\tau$ & \hspace{2mm} $\frac{1}{6}$ \hspace{5mm} $0$ \hspace{5mm} $0$ \hspace{1mm} \\ 
  $\nu_2$-Trimaximal$^*$ \rule{0mm}{4mm} & \hspace{1mm} $0$ \hspace{5mm} - \hspace{5mm} $0$ \hspace{5mm} - \hspace{1mm}
                & Democracy & \hspace{2mm} - \hspace{6mm} $0$ \hspace{5mm} - \hspace{1mm} \\ 
  Two-equal P-rows$^*$ \rule{0mm}{4mm}& \hspace{1mm} $0$ \hspace{5mm} - \hspace{5mm} - \hspace{5mm} $0$ \hspace{1mm}
                & e.g.\ $\mu$-$\tau$ & \hspace{2mm} - \hspace{6mm} $0$ \hspace{5mm} - \hspace{1mm} \\ 
  Two-equal P-columns \rule{0mm}{4mm}& \hspace{1mm} $0$ \hspace{5mm} - \hspace{5mm} - \hspace{5mm} $0$ \hspace{1mm}
                & e.g.\ $1$-$2$ & \hspace{2mm} - \hspace{6.5mm} - \hspace{5mm} $0$ \hspace{1mm} \\ 
  Altaerlli-Feruglio$^*$ \rule{0mm}{4mm}& \hspace{1mm} $0$ \hspace{5mm} - \hspace{2mm} $\frac{6G-1}{8}$ \hspace{2mm} $0$ \hspace{1mm}
                & e.g.\ $\mu$-$\tau$, $CP$ & \hspace{2mm} $0$ \hspace{5mm} $0$ \hspace{5mm} -  \hspace{1mm} \\ 
  Tri$\chi$maximal Mix.$^*$ \rule{0mm}{4mm} & \hspace{1mm} $0$ \hspace{5mm} - \hspace{5mm} $0$ \hspace{5mm} $0$ \hspace{1mm}
                & Dem., $\mu$-$\tau$ & \hspace{2mm} - \hspace{5mm} $0$ \hspace{5mm} - \hspace{1mm} \\ 
  Tri$\phi$maximal Mix.$^*$ \rule{0mm}{4mm}& \hspace{1mm} $0$ \hspace{5mm} $\frac{1}{6}$ \hspace{5mm} $0$ \hspace{5mm} - \hspace{1mm}
      & Dem., $CP$ & \hspace{2mm} $0$ \hspace{5mm} $0$ \hspace{5mm} - \hspace{1mm} \\ 
  Bimaximal Mix. \rule{0mm}{4mm}& \hspace{1mm} $0$ \hspace{5mm} $\frac{1}{8}$ \hspace{1mm} $-\frac{1}{32}$ \hspace{3mm} $0$ \hspace{1mm}
   \raisebox{-3mm}{\rule{0mm}{4mm}} & Dem., $CP$, $1$-$2$ & \hspace{2mm} $0$ \hspace{5mm} $0$ \hspace{5mm} $0$ \hspace{1mm}  \\ 
  \hline\hline
\end{tabular} 
\caption{Possible and proposed mixing schemes/constraints 
expressed in terms of our Flavour-Symmetric Mixing Observacles (FSMOs).
Those marked with an asterisk are still phenomenologically viable.
Although our four $L \leftrightarrow N$ symmetric variables ${\cal F}$, ${\cal G}$, ${\cal C}$ and ${\cal A}$ 
are sufficient to completely fix the mixing (up to permutations) 
we include ${\cal B}$ and ${\cal D}$ (and ${\cal J}^2$) for completeness.}
\end{table}

Some rather less restrictive flavour-symmetric 
constraints may also be written:
\begin{eqnarray}
8{\cal C}^3-27{\cal F}^2({\cal CG}-{\cal AF})=0 & \hspace{2mm} \Leftrightarrow \hspace{2mm} & 
            |U_{\alpha i}|^2=1/3 \hspace{8mm} ({\rm any} \, \alpha, {\rm any} \, i) \hspace{2mm} \label{c1}  \\
8{\cal B}^3-27{\cal F}^2({\cal BG}-{\cal DF})=0 & \hspace{2mm} \Leftrightarrow \hspace{2mm} & 
          |U_{\alpha i}|^2=|U_{\beta i}|^2 \hspace{5mm} (\alpha \ne \beta, {\rm any} \, i) \hspace{2mm} \label{c2} \\
8{\cal D}^3-27{\cal F}^2({\cal DG}-{\cal BF})=0 & \hspace{2mm} \Leftrightarrow \hspace{2mm} & 
          |U_{\alpha i}|^2=|U_{\alpha j}|^2 \hspace{5mm} ({\rm any} \, \alpha,i \ne j) \hspace{2mm} \label{c3}
\end{eqnarray}
corresponding respectively to one mixing-element with modulus-squared $1/3$ (Eq.~\ref{c1}),
two elements with equal modulus in the same column (Eq.~\ref{c2})
or, indeed, with equal modulus in the same row (Eq.~\ref{c3}).
A single mixing element zero corresponds to:
\begin{equation}
{\cal K}
=0 \hspace{2mm} {\rm and} \hspace{2mm}  {\cal J}=0 
\hspace{4mm} \Leftrightarrow \hspace{4mm} |U_{\alpha u}|=0 
\hspace{8mm} ({\rm any} \, \alpha, {\rm any} \, i) \label{k=0j=0}
\end{equation}
where:
\begin{equation}
{\cal K}:={\rm Det} \, K={\cal A}/27+{\cal F}^3/54-{\cal FC}/27-{\cal F}/54.
\end{equation}
An approach to Eq.~\ref{k=0j=0} such that
${\cal J} \rightarrow 0$ with 
${\cal K}\equiv 0$
offers an
explanation\cite{hds2009}\cite{hrs2007}\cite{mori}\cite{panic}
for one near-right unitarity-triangle angle, 
e.g.\ $\alpha \simeq 90^o$ in the quark~sector\cite{fz1995}.

One extraordinarily simple/powerful condition 
(now excluded even for the quarks):
\begin{equation}
{\cal B}={\cal D}  \hspace{2mm} \Leftrightarrow \hspace{2mm} 
           (|U|)=(|U^T|) \hspace{5mm} \hspace{2mm} \label{b=d}
\end{equation}
would have guaranteed a completely symmetric mixing (moduli) matrix!

While we have not succeeded to solve Eqs.~\ref{g=wxyz}-\ref{a=wxyz}
for $w,x,y,z$ in terms of ${\cal G,F,C,A}$,
we may understand rather simply how each of our observables
affects the symmetry of the $P$-matrix
by setting ${\cal F}=f$, ${\cal C}=c$ and ${\cal A}=a$ 
(where $f,c,a \ll {\cal G}$, \hspace{1mm} $0 < {\cal G} < 1/6$)
and then solving to leading order in each of $f,c$ and $a$:
\begin{equation}
P \sim \left( \matrix{ 
\rule{0pt}{13pt} \frac{1}{3}+2\sqrt{\frac{{\cal G}}{6}} +\frac{2c}{9{\cal G}}       
                          & \frac{1}{3} -\frac{4c}{9{\cal G}} & 
                                         \frac{1}{3}-2\sqrt{\frac{{\cal G}}{6}}+\frac{2c}{9{\cal G}}  \cr
  \rule{0pt}{13pt} \frac{1}{3}-\sqrt{\frac{{\cal G}}{6}}-\frac{f}{2\sqrt{6{\cal G}}} -\frac{c}{9{\cal G}} +\frac{a}{3{\cal G}}
                             & \frac{1}{3} +\frac{f}{\sqrt{6{\cal G}}} +\frac{2c}{9{\cal G}} & 
      \frac{1}{3} +\sqrt{\frac{{\cal G}}{6}}-\frac{f}{2\sqrt{6{\cal G}}} -\frac{c}{9{\cal G}} -\frac{a}{3{\cal G}} \cr
  \rule{0pt}{13pt} \frac{1}{3}-\sqrt{\frac{{\cal G}}{6}}+\frac{f}{2\sqrt{6{\cal G}}} -\frac{c}{9{\cal G}} -\frac{a}{3{\cal G}}
                     & \frac{1}{3} -\frac{f}{\sqrt{6{\cal G}}} +\frac{2c}{9{\cal G}} & 
            \frac{1}{3} +\sqrt{\frac{{\cal G}}{6}}+\frac{f}{2\sqrt{6{\cal G}}} -\frac{c}{9{\cal G}} +\frac{a}{3{\cal G}} } \right)
\end{equation}
An expansion about tribimaximal mixing is now readily achieved 
setting ${\cal G}=1/6-g$ and expanding to first order in $g$ also, with $g \ll 1/6$.
\begin{equation}
P(g,f,c,a) \sim \left( \matrix{ \rule{0pt}{13pt} \frac{2}{3}-g +\frac{4c}{3}       
                                                       & \frac{1}{3} -\frac{8c}{3} & 
                                                                      g+\frac{4c}{3}  \cr
  \rule{0pt}{13pt} \frac{1}{6}+\frac{g}{2}-\frac{f}{2}-\frac{2c}{3}+2a 
                             & \frac{1}{3} +f +\frac{4c}{3} & 
      \frac{1}{2} -\frac{g}{2}-\frac{f}{2}-\frac{2c}{3}-2a \cr
  \rule{0pt}{13pt} \frac{1}{6}+\frac{g}{2}+\frac{f}{2}-\frac{2c}{3}-2a
                     & \frac{1}{3} -f +\frac{4c}{3} & 
            \frac{1}{2} -\frac{g}{2}+\frac{f}{2}-\frac{2c}{3}+2a  } \right),
\end{equation}
thus forming the basis of an interesting
and potentially useful parameterisation 
of the mixing matrix in terms of small deviations\cite{park}\cite{pfh}\cite{king} 
from tribimaximal mixing.

\section{Extremisation of Flavour-Symmeric ``Actions/Potentials''}

To the extent that the neutrino mixing is in any sense ``maximal''
one might say that experiment {\em suggests} 
that some kind of ``extremisation'' may be at work here. 
Given Eq.~\ref{fgcatbm} above,
it is rather obvious
that (up to permutations) tribimaximal mixing
can be guaranteed by requiring that the flavour-symmetric function:
\begin{equation}
V({\cal G},{\cal F},{\cal C},{\cal A})={\cal F}^2+{\cal C}^2+{\cal A}^2+{\cal J}^2 \label{f2g2c2a2}
\end{equation}
be extremal with respect to ${\cal F},{\cal C},{\cal A},{\cal J}$, 
since $(\partial V/\partial {\cal F})_{CAJ}=2{\cal F}=0 \Rightarrow {\cal F}=0$~etc.
The choice of extremisation variables is not crucial here,
since such functions would generally yield TBM for arbitrary (e.g. PDG) choice of variables
(clearly the function Eq.~\ref{f2g2c2a2} is far from unique, 
e.g.\ arbitrary coefficients could obviously be included).

In our 1994 paper\cite{hs1994}, 
we tentatively floated the idea that extremisation 
might also account for the observed spectrum of masses.
The simplest example we had in mind was very trivial.
Take the action/potential 
to be just the determinant of the mass matrix ${\rm Det} \, M$
(we are considering a generic fermion type here,
and we are still thinking of $M=L,N,U\dots$ 
most naturally as  
a `hermitised' mass matrix as in Eq.~\ref{L=N=} above): 
\begin{equation}
V(M)={\rm Det} \, M = x y z (\sqrt{2} \Phi)^3
\end{equation}
where $x,y,z$ are now (c.f. Section~3) 
the usual (diagonalised) Yukawa couplings.
Taking the SM higgs field $\Phi$ to be essentially fixed
by more significant terms elsewhere in the Lagrangian,
we extremise with respect to the Yukawa couplings $x,y,z$~themselves: 
\begin{equation}
\left.
\begin{array}{l}
(\partial V/ \partial x)_{yz} = yz =0 \\
(\partial V/ \partial y)_{zx} = zx =0 \\
(\partial V/ \partial z)_{xy} = xy =0 
\end{array} \right\} \hspace{4mm} \Rightarrow \hspace{2mm} {\rm e.g.}
\left\{
\begin{array}{l}
x \ne 0 \\
y =0 \\
z = 0 
\end{array}
\right.
\hspace{3mm} {\rm or} \hspace{1mm}
\left\{
\begin{array}{l}
x = 0 \\
y \ne 0 \\
z = 0 
\end{array}
\right.
\hspace{3mm} {\rm or} \hspace{1mm}
\left\{
\begin{array}{l}
x = 0 \\
y =0 \\
z \ne 0 
\end{array}
\right.
\end{equation}
The solutions require two light (zero mass) fermions 
and one (potentially) heavy, i.e.\ non-zero mass fermion,
surely a good starting point as regards 
the observed quark and lepton mass spectra. 
Notice that the symmetry of our ``action'' forbids to tell which fermion 
will turn out to be heavy (indeed there is no distinction between $x,y,z$ a priori).
One might say that the choice, e.g.\ $z \ne 0$, is made spontaneously.

Our 2005 paper\cite{ext2005}
made a somewhat more serious attack 
on these kinds of issues, especially as regards the mixing,
adopting a strictly (Jarlskog) covariant approach.
We made particular use of a theorem in matrix calculus\cite{adler}:
\begin{equation}
\partial_X \; {\rm Tr} \; XA    
 \, = \, A^T \label{theo1}
\end{equation}
which enabled us to extremise directly 
with respect to the mass matrices 
(i.e.\ with respect to the Yukawa couplings, within the SM).
As a warm-up exercise, we began by extremising
the Jarlskog determinant ${\rm Det} \, C_{} = {\rm Tr } C_{}^3/3$ 
(here $C:=C_{11}:=-i[L,N]$): 
\begin{equation}
\begin{array}{l}
(\partial_L \, {\rm Tr} \; C_{11}^3/3)^T = +i[N,C_{}^2] = 0 \label{exsl} \\
(\partial_N \, {\rm Tr} \; C_{11}^3/3)^T = -i[L,C_{}^2] = 0 \label{exsn}
\end{array} \hspace{2mm}
\Rightarrow \hspace{2mm} 
P=\left( \matrix {  1/3 & 1/3 & 1/3 \cr
                 1/3 & 1/3 & 1/3 \cr
                 1/3 & 1/3 & 1/3 } \right) \hspace{2mm} \label{dc3}
\end{equation}
which, as expected, yields trimaximal mixing
(it should be said that $2 \times 2$ maximal mixing in any sector 
also provides a solution). 
When extremising for the mixing at fixed masses, 
the zeros on the RHS of Eqs.~\ref{dc3} get replaced
by polynomials in the mass matrices determining Lagrange multipliers
(having almost no effect in practice).

Similarly, extremising the sum of the Principal minors 
(${\rm Pr} \, C_{}=C_{}^2/2$):
\begin{equation}
\begin{array}{l}
(-\partial_L {\rm Tr} \, C^2/2)^T = +i[N,C]  =  \; 0  \label{expl} \\
(-\partial_N {\rm Tr} \, C^2/2)^T = -i[L,C]  =  \; 0 \label{expn} 
\end{array}\hspace{2mm}
\Rightarrow \hspace{2mm} {\rm e.g.} \hspace{2mm}
P=\left( \matrix {  1 & 0 & 0 \cr
                 0 & 1/2 & 1/2 \cr
                 0 & 1/2 & 1/2 } \right) \hspace{2mm}
\end{equation}
yields, on the same basis, 
uniquely $2 \times 2$ maximal mixing, as one might expect
(we are considering solutions 
which apply independently
of the values of the masses -
it turns out that there is also a unique 
mass-dependent solution\cite{ext2005},
which although not entirely without interest,
certainly does not agree with experiment). 
We remark that we see our extremisation equations,
Eqs.~\ref{expl}, 
as analogous to the Yang-Mills \cite{yngm} equations,
cf.\ $[\nabla_{\mu}, F^{\mu \nu}]=0$,
derivable from 
the quadratic Lagrangian ${\cal L} = -{\rm Tr} \; F^2/2$.
Yukawa couplings are thus seen as dynamical variables,
analogous to gauge fields.

\subsection{Extremisation of an Arbitrary Function of the Jarlskog Commutator}

We can consider extremising some function of both Tr $C^2$ and Tr $C^3$ {\em together}
to obtain potentially more realistic predictions
(note that ${\rm Tr} \, C =0$ identically for a commutator).
If the function to be extremised is denoted $A$
then, at the extremum:
\begin{eqnarray}
r \; \partial_L {\rm Tr} C^2+\partial_L {\rm Tr} C^3 =0  \label{rdifl} \label{dlc2c3}\\
r \; \partial_N {\rm Tr} C^2+\partial_N {\rm Tr} C^3 =0  \label{rdifn} \label{dnc2c3}
\end{eqnarray}
is the most general matrix extremisation condition
(traces of higher powers than third are always reducible to cubic or less via the characteristic equation). 
In Eqs.~\ref{dlc2c3}-\ref{dnc2c3}
the dimensionful scalar parameter $r$ 
is just the ratio of partial derivatives:
\begin{equation}
r=\frac{\partial A \, / \, \partial {\rm Tr} C^2}{\partial A \, / \, \partial {\rm Tr} C^3} \label{rdef}
\end{equation}
likewise understood as being evaluated at the extremum.
Essentially, the resulting constraint equations
are simply a linear combination of the constraints Eqs.~\ref{exsl}-\ref{expn}.

Seeking solutions, we may work in any basis we choose,
and we choose to work in the usual charged-lepton flavour basis 
where the charged-lepton mass matrix is diagonal,
in a phase convention where all the imaginary parts 
of the off-diagonal elements in the neutrino mass-matrix  
are equal up to signs 
(so that the matrix of imaginary parts
is proportional to the epsilon matrix~$\epsilon$, 
hence `the epsilon phase convention' \cite{sim2004}).
The neutrino mass-matrix is then 
(redefining $x,y,z$ yet again!):
\begin{eqnarray}
     \matrix{  \hspace{0.1cm} e \hspace{0.5cm}
               & \hspace{0.5cm} \mu \hspace{0.5cm}
               & \hspace{0.5cm} \tau  \hspace{0.1cm} }
                                      \hspace{5.7cm} \nonumber \\
N=M_{\nu} =
\matrix{ e \hspace{0.0cm} \cr
         \mu \hspace{0.0cm} \cr
         \tau \hspace{0.0cm} } 
\left( \matrix{ a  &
                      z+id &
                              y-id \cr
                z-id &
                    b &
                             x+id \cr
      \hspace{2mm} y+id \hspace{2mm} &
         \hspace{2mm}  x-id \hspace{2mm} &
           \hspace{2mm} c  \hspace{2mm} \cr } \right) \hspace{0.8cm}
\epsilon= \left( \matrix{ 0 & 1 & -1\cr
                         -1 & 0 & 1 \cr
      1 & -1 & 0 \cr } \right) \label{eps}
\end{eqnarray}
The seven variables $a$, $b$, $c$, $x$, $y$, $z$, $d$ then determine 
the three neutrino masses and the four mixing parameters
(with unphysical phase dependence neatly eliminated!).\footnote{
In this section we continue to consider hermitian mass matrices,
in the first instance (rather than, say, hermitian squares),
as being the more fundamental and natural for extremisation.}

In solving for fixed masses,
there is also a determinantal consistency condition: 
\begin{equation}
\left| \matrix{ \hspace{2mm} 1 \hspace{2mm} & 
        \hspace{2mm}a \hspace{2mm} & \hspace{2mm}a^2+y^2+z^2+2d^2 \hspace{2mm} \cr
               1 & b & x^2+b^2+z^2+2d^2 \cr
               1 & c & x^2+y^2+c^2+2d^2} \right| = 0. \label{detc}
\end{equation}
formed from the cefficients of the relevant Lagrange multipliers.
Looking initially for solutions with `democracy' symmetry 
(as solves Eq.~\ref{detc}) we have:
\begin{equation}
\left.
\begin{array}{l}
a-b=x-y \\
b-c=y-z \\
c-a=z-x 
\end{array} \right\} \hspace{4mm} 
\Rightarrow \hspace{2mm} 
{\rm `democracy'} \hspace{2mm}
\left\{
\begin{array}{l}
a = x + \sigma \\
b = y + \sigma  \\
c = z + \sigma \hspace{5mm} \label{s3dem}
\end{array}
\right.
\end{equation}
where $\sigma$ is a constant mass 
offset.

Since $L$ is diagonal, the off-diagonal elements 
in Eq.~\ref{rdifl} vanish.
The real parts~give:
\begin{eqnarray}
 \hspace{2.0cm} 2 \; (m_{\mu}+m_{\tau}-2m_e) \, (d^2-y \, z) \; +
         \; 2\, (m_{\mu}-m_{\tau}) \, (y-z) \, x \; = \; 0 \hspace{1.6cm} \nonumber \\
 \hspace{2.0cm} 2 \; (m_{\tau}+m_e-2m_{\mu}) \, (d^2-z \, x) \; +
         \; 2\, (m_{\tau}-m_e) \, (z-x) \, y \; = \; 0 \hspace{1.6cm} \label{rsr}  \\
 \hspace{2.0cm} 2 \; (m_e+m_{\mu}-2m_{\tau}) \, (d^2-x \, y) \; +
         \; 2\, (m_e-m_{\mu}) \, (x-y) \, z \; = \; 0 \hspace{1.6cm} \nonumber
\end{eqnarray}
explicitly cyclically symmetric, and the imaginary parts give:
\begin{eqnarray}
  2 r \, d \, (m_{\mu}+m_{\tau}-2m_e)  (y & \hspace{-3.5mm} + \hspace{-3.5mm} & z) \;   + 
          \; 2 r \, d \, (m_{\mu}-m_{\tau}) \, (y-z) \; \nonumber \\
   \;  = & \hspace{-5mm} & \;\hspace{-5mm}  3 \, (m_e-m_{\mu}) \, (m_{\tau}-m_e) \; 
   (y-z)\,(d^2-xy-yz-zx)  \hspace{0.5cm} \nonumber \\
  2 r \, d \, (m_{\tau}+m_e-2m_{\mu})   (z & \hspace{-3.5mm} + \hspace{-3.5mm} & x) \;  + 
          \; 2 r \, d \, (m_{\tau}-m_e) \, (z-x) \;  \nonumber \\
  \;  =  & & \hspace{-5mm} \; 3 \, (m_{\mu}-m_{\tau}) \, (m_e-m_{\mu}) \; 
   (z-x)\,(d^2-xy-yz-zx) \hspace{0.5cm} \label{rsi} \\
  2 r \, d \, (m_e+m_{\mu}-2m_{\tau})  (x & \hspace{-3.5mm} + \hspace{-3.5mm} & y) \;   + 
          \; 2 r \, d \, (m_e-m_{\mu})  (x-y) \;  \nonumber \\
     \;  =&  & \hspace{-5mm} \; 3 \, (m_{\tau}-m_e) \, (m_{\mu}-m_{\tau}) \; 
   (x-y)\,(d^2-xy-yz-zx) \nonumber \hspace{0.5cm}
\end{eqnarray}
also explicitly cyclic symmetric.
A solution to Eq.~\ref{rsr}-\ref{rsi} is:
\begin{eqnarray}
x = \pm \frac{\sqrt{XYZ}}{X} \hspace{2.0cm} X=d^2-\frac{2dr(m_{\mu}-m_{\tau})}
{3(m_e-m_{\mu})(m_{\tau}-m_e)} \hspace{2mm} \nonumber \\
y = \pm \frac{\sqrt{XYZ}}{Y} \hspace{2.0cm} Y=d^2-\frac{2dr(m_{\tau}-m_e)}
{3(m_{\mu}-m_{\tau})(m_e-m{\mu})} \label{solrs} \\
z = \pm \frac{\sqrt{XYZ}}{Z} \hspace{2.0cm} Z=d^2-\frac{2dr(m_e-m_{\mu})}
{3(m_{\tau}-m_e)(m_{\mu}-m_{\tau})} \hspace{2mm} \nonumber
\end{eqnarray}
where the cyclic symmetry is clearly respected in the solution.

The operative parameter is now $r/d$,
fixing the mixing and (if we are prepared to assume,
eg.\ $m_1 << m_2 << m_3$) also the neutrino mass hierarchy.
Clearly, from Eq.~\ref{solrs} the mixing approaches the 
so-called `simplest'\cite{sim2004} form ($|y|,|z| << |x|$)
as the denominator factor $X$ goes to zero
($X \rightarrow 0$ corresponds to $r/d \rightarrow 0.168$ GeV). 
In the extreme limit, the mixing matrix
tends finally to the familiar tri-bimaximal \cite{hps2002} form
(with an exact degeneracy $\Delta m_{12}^2 \rightarrow 0$ occuring at the pole).
This qualitative tendency to the `simplest' anatz\cite{sim2004} seen here
(as the approach to tribimaximal mixing) 
is the most phenomenologically encouraging result
we have found so far from this part of our extremisation programme.
In quantitative terms, however,
even this tendency brings little satisfaction in practice. 

Constraining $r/d$ to the observed mass hierarchy (assuming $m_1 << m_2 << m_3$) we have
$\Delta m_{12}^2/\Delta m_{23}^2 \simeq m_2^2/m_3^2 \simeq 0.035$ for $r/d= 0.245$ GeV,
with a mixing matrix:
\begin{eqnarray}
     \matrix{  \hspace{0.1cm} \nu_1 \hspace{0.6cm}
               & \hspace{0.5cm} \nu_2 \hspace{0.6cm}
               & \hspace{0.5cm} \nu_3  \hspace{0.2cm} }
                                      \hspace{2.8cm} 
          \matrix{  \hspace{0.1cm} \nu_1 \hspace{0.2cm}
               & \hspace{0.4cm} \nu_2 \hspace{0.2cm}
               & \hspace{0.4cm} \nu_3  \hspace{0.2cm} }
                                      \hspace{0.6cm}\nonumber \\
(|U_{l \nu}|^2) \; = \;
\matrix{ e \hspace{0.2cm} \cr
         \mu \hspace{0.2cm} \cr
         \tau \hspace{0.0cm} } \hspace{-2mm}
\left( \matrix{ .48072  &
                      .33333 &
                              .18595 \cr
                ..40735 &
                    .33333 &
                             .25932 \cr
      \hspace{2mm} .11194 \hspace{2mm} &
         \hspace{2mm}  .33333 \hspace{2mm} &
           \hspace{2mm} .55473  \hspace{2mm} \cr } \right)
\; \not\simeq \;
\matrix{ e \hspace{0.2cm} \cr
         \mu \hspace{0.2cm} \cr
         \tau \hspace{0.0cm} } \hspace{-2mm}
\left( \matrix{ 2/3  &
                      1/3 &
                              0 \cr
                1/6 &
                    1/3 &
                             1/2  \cr
      \hspace{2mm} 1/6 \hspace{2mm} &
         \hspace{2mm}  1/3 \hspace{2mm} &
           \hspace{2mm} 1/2  \hspace{2mm} \cr } \right) \label{ntbm} \hspace{2mm}
\end{eqnarray}
We see (disappointingly)
that in practice we are too far from the pole for
the `simplest' \cite{sim2004} approximation to apply, whereby,
we are left with too large a value for $|U_{e3}|$ above.
Note that the poles at $Y \rightarrow 0$ ($r/d \rightarrow 0.148$ GeV) 
and $Z \rightarrow 0$ ($r/d \rightarrow 42.4$ GeV)
correspond to `permuted' forms of tribimaximal mixing
having a $\tau-e$ symmetry (with $|U_{\mu 3}| \rightarrow 0$) 
and an $e-\mu$ symmetry (with $|U_{\tau 3}| \rightarrow 0$) respectively,
and do not therefore seem to be phenomenologically relevant.

Although democracy (Eq.~\ref{s3dem}) 
is an elegant
and promising way to implement 
the determinant condition (Eq.~\ref{detc}),
it will be clear that it is not forced on us here.
We therefore explore the trajectory of solutions generated 
dropping this constraint,
as we move away from the solution Eq.~\ref{ntbm}
keeping the neutrino mass hierachy constant
(ie.\ taking $\Delta m_{12}^2/\Delta m_{23}^2 \simeq m_2^2/m_3^2 \simeq 0.035$).
We find that we can reduce the value of $|U_{e3}|$,
but that $|U_{e2}|$ also decreases 
(and furthermore decreases faster than $|U_{e3}|$):
\begin{eqnarray}
     \matrix{  \hspace{0.1cm} \nu_1 \hspace{0.6cm}
               & \hspace{0.4cm} \nu_2 \hspace{0.6cm}
               & \hspace{0.4cm} \nu_3  \hspace{0.2cm} }
                                      \hspace{2.8cm} 
          \matrix{  \hspace{0.1cm} \nu_1 \hspace{0.2cm}
               & \hspace{0.4cm} \nu_2 \hspace{0.2cm}
               & \hspace{0.4cm} \nu_3  \hspace{0.2cm} }
                                      \hspace{0.6cm}\nonumber \\
(|U_{l \nu}|^2) \; = \;
\matrix{ e \hspace{0.2cm} \cr
         \mu \hspace{0.2cm} \cr
         \tau \hspace{0.0cm} } \hspace{-2mm}
\left( \matrix{ .64243  &
                      .21957 &
                              .13800 \cr
                .30883 &
                    .34235 &
                             .34882 \cr
   \hspace{2mm} .04874 \hspace{2mm} &
         \hspace{2mm}  .43808 \hspace{2mm} &
           \hspace{2mm} .51318  \hspace{2mm} \cr } \right)
\; \not\simeq \;
\matrix{ e \hspace{0.2cm} \cr
         \mu \hspace{0.2cm} \cr
         \tau \hspace{0.0cm} } \hspace{-2mm}
\left( \matrix{ 2/3  &
                      1/3 &
                              0 \cr
                1/6 &
                    1/3 &
                             1/2  \cr
      \hspace{2mm} 1/6 \hspace{2mm} &
         \hspace{2mm}  1/3 \hspace{2mm} &
           \hspace{2mm} 1/2  \hspace{2mm} \cr } \right) \label{nntbm} \hspace{2mm}
\end{eqnarray}   
Eq.~\ref{nntbm} is simply a representative `compromise' solution
which (even so) is still very far from viable.
The mass matrix parameters are 
$(x/d,y/d,z/d)=(-3.270,-1.883,2.240)$; $(a/d,b/d,c/d)=(1.725,3.979,5.735)$ 
and $r/d=1.086$ GeV.
We have not suceeded to obtain analytical solutions
when dropping the democracy constraint.

\section{Our Best Guess So Far: The `Simplest' Ansatz}

We have seen that extremisation 
of a general flavour-symmetric scalar function 
of the Jarlskog commutator $C$,
may be said to {\em point to} the `simplest' ansatz. 
The `simplest' ansatz 
was put forward on aesthetic grounds in 2004\cite{sim2004},
having been previously proposed
as ``an amusing specialisation of tri-$\chi$-maximal mixng''
in 2002\cite{hs2002}.

To appreciate the `simplest' ansatz\cite{sim2004},
one had best focus again on 
hermitian-squares of mass-matrices.
In the usual charged-lepton flavour basis 
(keeping the epsilon phase convention), 
one simply takes the neutrino mass matrix (hermitian-square)
to be a linear combination
of the $3 \times 3$ identity matrix, 
the $\mu-\tau$-exchange operator $E$ (Eq.~\ref{mtop})(with a negative coefficient)
and the epsilon matrix $\epsilon$ (Eq.~\ref{eps}):
\begin{eqnarray}
M_{\nu}M_{\nu}^{\dagger} & = & \sigma I+xE+id\epsilon \\
 & = & \left( \matrix{ \sigma +x & id & -id \cr
                        -id & \sigma &  x+id \cr
                         id & x-id &  \sigma \cr } \right) \label{msimp}
\end{eqnarray}
with the two mass-squared differences fixing the parameters $x$ and $d$ ($x < 0$):
\begin{eqnarray}
\Delta m^2_{\rm atm} = m_3^2-m_1^2 & = & 2\sqrt{x^2+3d^2} \\
\Delta m^2_{\rm sol} = m_2^2-m_1^2 & = & \sqrt{x^2+3d^2}+x .
\end{eqnarray}
The parameter $\sigma$ in Eq.~\ref{msimp}
represents an overall offset 
on the neutrino mass-squared spectrum,
precisely determined only when the masses of all three neutrinos 
(i.e.\ including the lightest neutrino) are known.
The offset $\sigma$ cannot influence the mixing.

Clearly, the mass matrix (Eq.~\ref{msimp}) 
commutes with the democracy operator (Eq.~\ref{demsym})
and with the $\mu-\tau$-reflection operator (Eq.~\ref{mtsym}),
so that the `simplest' ansatz builds-in the
`democracy' and `mutativity' symmetries.
The resulting mixing matrix 
is simply a reparametrisation of tri$\chi$maximal mixing\cite{hs2002}, 
but with the mixing angle determined:
\begin{equation}
U=\left(\matrix{ \sqrt{\frac{2}{3}}c_{\chi}
                & \sqrt{\frac{1}{3}} & -i\sqrt{\frac{2}{3}}s_{\chi} \cr
-\frac{c_{\chi}}{\sqrt{6}}+i\frac{s_{\chi}}{\sqrt{2}} 
                & \sqrt{\frac{1}{3}}
                    & -\frac{c_{\chi}}{\sqrt{2}}+i\frac{s_{\chi}}{\sqrt{6}} \cr
     -\frac{c_{\chi}}{\sqrt{6}}-i\frac{s_{\chi}}{\sqrt{2}}
             & \sqrt{\frac{1}{3}}
                   &  \frac{c_{\chi}}{\sqrt{2}}+i\frac{s_{\chi}}{\sqrt{6}}  }
\right) \hspace{12mm} \tan 2 \chi= \sqrt{3}d/x .
\end{equation}
This leads us finally to a simple prediction
for the reactor neutrino mixing angle
in terms of measured mass-squared differences:
\begin{eqnarray}
\sin \theta_{13}=\sqrt{2\Delta m^2_{\rm atm}/3\Delta m^2_{\rm sol}} \simeq 0.13 \pm 0.03 .
\end{eqnarray}
This long-standing prediction\cite{hs2002}\cite{sim2004} 
lies in the peak
of Carl Albright's distribution!\cite{carl}

\section{Acknowledgements}
PFH acknowledges the financial support of the European Community 
under the European Commission Framework Programme 7 
Design Study: EUROnu, Project Number 212372 
and the CfFP Rutherford Appleton Laboratory.
WGS presented very similar talks 
at the UK HEP-Forum on ``Neutrino Physics'', 
Abingdon, UK, 17-18 April 2008,
the Symposium on ``Physics of Massive Neutrinos'' (PMN08) 
Milos, Greece  20-22 May  2008 
and at the Topical Conference on 
``Elementary Particles, Astrophysics and Cosmology''
(Miami-2008)
Fort Lauderdale, Florida, USA. 16 - 21 Dec 2008.
WGS wouild like to thank the European Network 
of Theoretical Astroparticle Physics ILIAS/N6 
under contract number RII3-CT-2004-506222 
for support at the Symposium in MILOS,
and most importantly also Milla Baldo-Ceolin for her~patience.


\newpage

\begin{thebibliography}{99}
\bibitem{hps1999} P. F. Harrison, D. H. Perkins and W. G. Scott. 
Phys. Lett. B 458 (1999) 79. hep-ph/9904297
\bibitem{wolf1978} L. Wolfenstein Phys. Rev. D18 (1978) 958.
\bibitem{cab1978} N.\ Cabibbo Phy. Lett. B 72 (1978) 333. 
\bibitem{hs1994} P. F. Harrison, W. G. Scott. 
Phys. Lett. B333 (1994) 471. hep-ph/9406351.
\bibitem{hps1995} P. F. Harrison, D. H. Perkins and W. G. Scott. 
Phys. Lett. B349 (1995) 137. \\
http://ccdb4fs.kek.jp/cgi-bin/img\raisebox{-1mm}{-}index?9503190
\bibitem{hps1996} P. F. Harrison, D. H. Perkins and W. G. Scott. 
Phys. Lett. B374 (1996) 111. hep-ph/9601346  
\bibitem{hps1997} P. F. Harrison, D. H. Perkins and W. G. Scott. 
Phys. Lett. B396 (1997) 186. hep-ph/9702243
\bibitem{mesh1990} P. Kaus and S. Meshkov, Mod Phys. Lett. A 3 (1988) 1251. \\
P. Kaus and S. Meshkov, Phys. Rev. D 42 (1990) 1863. \\
P. Kaus and S. Meshkov, Phys. Lett. B 611 (2005) 147. hep-ph/0410024.
\bibitem{pak1993} A. Acker et al.  
Phys. Lett. B298 (1993) 149. 
\bibitem{fz1996} H. Fritzsch and Z-z Xing, Phys. Lett. B 372 (1996) 265. hep-ph/9509389 
\bibitem{af1998} G. Altarelli and F. Feruglio, 
Phys.\ Lett.\ B 439 (1998) 112. hep-ph/9807353.
\bibitem{bmx1998} V. Barger et al. 
Phys. Lett. B437 (1998) 107 hep-ph/9806387.
\bibitem{gold} A. J. Bahz, A. S. Goldhaber, M. Goldhaber. 
Phys. Rev. Lett. 81 (1998)~5730.
\bibitem{glash} H. Giorgi and S. L. Glashow. hep-ph/9808293. F. Vissani hep-ph/9708483.
\bibitem{hps2002} P. F. Harrison, D. H. Perkins and W. G. Scott
Phys. Lett. B530 (2002) 167 hep-ph/0202074.
\bibitem{tdlvid} T. D. Lee, CERN lecture ``Symmetry and Asymmetry in EW Interaction'' Aug 2007. 
\hspace{0.1mm} See: \hspace{0.1mm} http://indico.cern.ch/conferenceDisplay.py?confId=19674  
\bibitem{ma2004} E. Ma Phys. Rev. D70 (2004) 031901. hep-ph/0404199. Also arXiv:0905.0221 
\bibitem{af2005} G. Altarelli and F. Feruglio, Nucl. Phys. B720 (2005) 64. hep-ph/0504165 \\ 
                 G. Altarelli 
et al. Nucl.\ Phys.\ B775 (2007) 31. hep-ph/0610165. \\
                 G. Altarelli, Neutrino Telescopes, Venice, Italy 2007 arXiv:0705.0860 [hep-ph]. \\
See also: hep-ph/0508053, hep-ph/0610164, hep-ph/0611117, arXiv:0711.0161.
\bibitem{zee2007} X. G. He and A. Zee Phys. Lett. B645 (2007) 427. hep-ph/0607163. \\
                  A. Zee, Phys.\ Lett.\ B 630 (2005) 58. hep-ph/0508278.
\bibitem{ross}  I. de Medeiros Varzielas, S. F. King, G. G. Ross, \\
Phys. Lett. B648 (2007) 201. arXiv:hep-ph/0607045 
\bibitem{fram}  P. H. Frampton, T. W. Kephart, S. Matsuzaki.\\
  Phys.\ Rev.\ D78 (2008) 073004. arXiv:0807.4713,  See also arXiv:0902.1140
\bibitem{litt} D. E. Littlewood,
The Theory of Group Characters\dots, 
OUP 1940. 
\bibitem{pass} D. S. Passman, The Algebraic Structure of Group Rings, Wiley (1977).
\bibitem{hs2002} P. F. Harrison, W. G. Scott. Phys. Lett. B 535 (2002) 163. hep-ph/0203209
\bibitem{hs2003} P. F. Harrison, W. G. Scott. Phys. Lett. B 557 (2003) 76. hep-ph/0302025
\bibitem{hsmt} P. F. Harrison, W. G. Scott. Phys. Lett. B 547 (2002) 219. hep-ph/0210197
\bibitem{ven2002} P. F. Harrison and W. G. Scott. NOVE-II, Venice 2003 hep-ph/0402006
\bibitem{sim2004} P. F. Harrison, W. G. Scott. Phys. Lett. B 594 (2004) 324. hep-ph/0403278
\bibitem{ext2005} P. F. Harrison, W. G. Scott. Phys. Lett. B 628 (2005) 93. hep-ph/0508012
\bibitem{lam2006} C. S. Lam, Phys. Lett. B640 (2006) 260. hep-ph/0606220 
\bibitem{bhs2006} J.D. Bjorken, P.F. Harrison, and W. G. Scott, Phys. Rev. D 74 (2006) 073012. hep-ph/0511201.
\bibitem{hds2009} P. F. Harrison, S. Dallison and W. G. Scott. arXiv:0904.3077
\bibitem{leecp} R. Friedberg and T. D. Lee, HEP \& NP 30 (2006) 591, hep-ph/0606071
\bibitem{leecpv} R. Friedberg and T. D. Lee, Annals Phys. 323 (2008) 1087.  arXiv:0705.4156.
\bibitem{lam2001} C. S. Lam, Phys. Lett. B 507 (2001) 214, hep-ph/0104116.
\bibitem{kita} T. Kitabayashi, M. Yasue. Phys.Lett. B621 (2005) 133. hep-ph/0504212
\bibitem{jarlinv} C. Jarlskog, Phys. Rev. Lett 55 (1985) 1039. ZPhys C 29 (1985) 491.
\bibitem{braninv} J. Bernabeu, G. C. Branco and M. Gronau, Phys. Lett B 169 (1986) 243. \\
G. C. Branco, L. Lavoura and M. N. Rebelo, Phys. Lett. B 180 (1986) 264.
\bibitem{hrs2007} P. F. Harrison, D. R. J. Roythorne and W. G. Scott. \\
Phys. Lett. B 657 (2007) 210. arXiv:0709.1439 
\bibitem{db} G. Kemper et al. Experimental Mathematics 10:4 (2001) 537.
\bibitem{pkq} P. F. Harrison, W. G. Scott and T. J. Weiler. Phys. Lett. B 641 (2006) 372. hep-ph/0607335
\bibitem{hawaii} P. F. Harrison and W. G. Scott 
``Generalisation of Tri-Bimaximal Mixing and \\ Flavout-Symmetric Constraints'',
presented by P. F. Harrison at the Joint Meeting of Pacific Region Particle Physics
Communities. Oct 29-Nov 3 (2006). \\
See: http://www.phys.hawaii.edu/indico/conferenceDisplay.py?confId=3
\bibitem{mori} P.F. Harrison, D.R.J. Roythorne, W.G. Scott,  
Proc.\ 43rd Rencontres de Moriond, 
EW Interactions, La Thuile, Italy (2008) arXiv:0805.3440 [hep-ph].
\bibitem{panic} P.\ F.\ Harrison, D.\ R.\ J.\ Roythorne and W.\ G.\ Scott,
Proc.\ 18th Particle and Nuclei Intl.\ Conf.\ (PANIC08),
Eilat, Israel (2008).  arXiv:0904.3014 [hep-ph].
\bibitem{fz1995} H.\ Fritzsch and Z.-z.\ Xing, Phys.\ Lett.\ B 353 (1995) 114. hep-ph/9502297.
\bibitem{park} S. Parke, Weak Ints. and Neutrinos (WIN05), Delphi, Greece. June (2005). \\
See: http://conferences.phys.uoa.gr/win05/PLENARY/parke.pdf
\bibitem{pfh} P. F. Harrison, 3rd Metting of International Scoping Study, RAL, April 2006. \\
              See: http://www.hep.ph.ic.ac.uk/uknfic/iss0406/talks/physics/
\bibitem{king} S. F. King, Phys. Lett. B659 (2008) 244, arXiv:0710.0530 [hep-ph]
\bibitem{adler} S. L. Adler, Quantum Theory as an Emergent Phenomenon, CUP (2004). \\
                P. Pearle, quant-ph/0602078. S. L. Adler, hep-th/9703053, hep-th/0510120.
\bibitem{yngm} R.\ L.\ Mills and C.\ N.\ Yang, Phys.\ Rev.\ D 96 (1954) 191. 
\bibitem{carl} C. H. Albright at this meeting, arXiv:0905.0146. See also arXiv:0803.4176.
\end{thebibliography}
\end{document}